\documentstyle [aps,epsfig]{revtex}
\catcode`\@=11
\newcommand{\fcaption}[1]{\vspace{1ex}
        \refstepcounter{figure}
        \setbox\@tempboxa = \hbox{\footnotesize {\bf ig.~\thefigure.} #1}
        \ifdim \wd\@tempboxa > 8cm
           {\begin{center}
        \parbox{8cm}{\footnotesize\baselineskip=8pt {\bf Fig.~\thefigure.} #1}

            \end{center}}
        \else
             {\begin{center}
             {\footnotesize {\bf Fig.~\thefigure.} #1}
              \end{center}}
        \fi}
\newcommand{\be}{\begin{equation}}
\newcommand{\ee}{\end{equation}}
\newcommand{\bea}{\begin{eqnarray}}
\newcommand{\eea}{\end{eqnarray}}

\def\pmb#1{\setbox0=\hbox{#1}
\kern-.025em\copy0\kern-\wd0
\kern.05em\copy0\kern-\wd0
\kern-.025em\raise.0433em\box0}
\newcommand{\BBU}{1\hspace*{-0.42ex}\rule{0.03ex}{1.48ex}\hspace*{0.44ex}}

\begin{document}

\title{Spin-stiffness and topological defects in two-dimensional frustrated spin systems}
\vspace{2cm}

\author{Michel Caffarel$~^{1}$, Patrick Azaria$~^{2}$, Bertrand Delamotte$~^{3}$,
and Dominique Mouhanna$~^{3}$}

\vspace{0.5cm}

\address{ $^1$  CNRS-Laboratoire de Chimie Th\'eorique,
Universit{\'e} Pierre et Marie Curie, 4 Place Jussieu, 75252 Paris,
France \\
$^2$ CNRS-Laboratoire de Physique Th\'eorique des Liquides,
Universit\'e Pierre et Marie Curie, 4 Place Jussieu, 75252 Paris,
France \\
$^3$ CNRS-Laboratoire de Physique Th\'eorique et Hautes Energies,
Universit\'e Pierre et Marie Curie, Universit\'e Denis Diderot, 
4 Place Jussieu, 75252 Paris, France }
\vspace{3cm}
 
\address{\rm (Received: )}
\address{\mbox{ }}
\address{ \parbox{14cm}{\rm \mbox{ }\mbox{ }
%\begin{abstract}
%\par
%%%%%%%
Using a {\it collective} Monte Carlo algorithm we study the low-temperature
and long-distance properties of two systems of two-dimensional classical tops. Both systems have the same
spin-wave dynamics (low-temperature behavior) as a large class of Heisenberg frustrated spin systems. 
They are constructed so that to differ only by their topological properties. The spin-stiffnesses
for the two systems of tops are calculated for different temperatures and 
different sizes of the sample.
This allows to investigate the role of topological defects in frustrated spin systems.
Comparisons with Renormalization Group results based on a Non Linear Sigma model
approach and with the predictions of some simple phenomenological model taking into account the topological
excitations are done.
}}
\address{ \mbox{ } }
\address{ \parbox{14cm} {\rm PACS No: 05.10.Ln, 75.10.Hk, 11.10.Hi, 11.10.Lm } }
\maketitle
\section{Introduction}
The long-distance behavior of the classical Heisenberg AntiFerromagnet on a Triangular lattice 
(HAFT model) has been the subject of much interest. In three dimensions a most important issue is the 
nature of the universality of its phase transition
$^{\cite{bailin,garel,kawamura1,aza1,kawamura10,tissier1,tissier2}}$. 
In two dimensions, this model has also been widely studied since it exhibits 
a non-trivial finite-temperature behavior due to the presence of topological excitations. 
Topology enters the problem since the order parameter of the model belongs to $SO(3)$ whose 
first homotopy group is $\pi_1[SO(3)]=Z_2$. 
As a consequence, there exist topologically stable point defects -- called vortices -- for this two-dimensional 
system. Arguments involving entropy and energy of the defects suggest the occurence of a change of behavior
at a finite temperature $T_V$ between a pure spin-wave regime with confined vortices for $T<T_V$ and
a regime of free vortices for $T>T_V$. Several Monte Carlo studies of the HAFT model 
$^{\cite{kawamura3,southern,southern1,wintel}}$
and of some generalizations with an easy-axis exchange anisotropy $^{\cite{stephan,stephan2,capriotti}}$
have indeed revealed the existence of various regimes resulting from the presence of defects.

Here, our purpose is to shed some light on the interplay between vortices and spin-waves
in 2D by studying  with Monte Carlo simulations two lattice models of ferromagnetically interacting tops. Both 
models have the {\it same} spin-wave dynamics as the original HAFT model but they differ by their topological 
properties: the first one has the same topological content as the HAFT model, 
the second one is topologically trivial. The role played by the topological defects emerges 
from the comparison between these two models. Note that this comparative study would have been 
more difficult to implement directly on the original HAFT model.
The physical quantity we consider in our study is the spin-stiffness which, for a spin system 
on a finite lattice of size $L$, measures the free-energy increment resulting from a twist of the boundary
conditions$^{\cite{chakravarty3,aza7}}$. 
The spin-wave part of the spin-stiffness identifies with the coupling constant of the Non 
Linear Sigma (NL$\sigma$) model renormalized at scale $L$ by thermal fluctuations. 
Accordingly, the behavior of the spin-stiffness as a function of the size of the lattice provides a direct 
test of the perturbative Renormalization Group (RG) predictions of the NL$\sigma$ model. 
For the models of tops studied here we expect that, at sufficiently 
low temperature --for $T$ significantly smaller than $T_V$--  where the physics is 
dominated by pure spin-waves, the behavior of the spin-stiffnesses agrees with the RG predictions. 
On the other hand, we also expect that, for a topologically non-trivial model, near $T_V$, 
the vortices also contribute to the spin-stiffness. 
For such a model the behavior of the spin-stiffness must disagree with the standard RG predictions. 
One great advantage of the spin-stiffness $\rho$ is that, in contrast with the correlation length $\xi$
which cannot be easily computed at very low temperature since it diverges typically as $\exp(1/T)$, 
$\rho$ has a smooth behavior at low temperature.
It is thus, in principle, easily computable.
Regarding Monte Carlo simulations, a central aspect is that, at the low temperatures we are interested in, 
the dynamics of 2D spin systems is governed by strongly correlated spin-waves, 
independently of the presence of vortices. These modes are responsible for a severe critical slowing down 
which makes difficult the convergence of simulations based on 
local algorithms in which one spin is flipped at each Monte Carlo step (``local update'' Monte Carlo schemes).
To resort to collective algorithms based on global updates (construction of clusters) 
is then important $^{\cite{swendsen,wolff}}$. 
However, as well-known, such algorithms work well for ferromagnetic systems but not for frustrated ones. 
Note that, together with the ability of comparing topologically different models, this aspect is an additional 
motivation to consider ferromagnetic top models rather than the original HAFT one. 
Actually, the implementation of the basic rules of collective algorithms for systems consisting of tops is 
itself not so clear. However, this difficulty can be circumvented by rewriting the models of tops 
considered here as ferromagnetic four-component spin systems 
while, of course,  preserving both their spin-wave and topological contents. 
Thanks to these various tricks and to the cluster algorithm we are then able to scan a large temperature 
range below $T_V$ while fully controlling the convergence of our simulations.
It should be noted that previous Monte Carlo calculations of the spin-stiffness 
on related models (HAFT model and generalizations) have been done by using Monte Carlo schemes 
with local updates$^{\cite{kawamura3,southern,southern1,wintel,stephan,stephan2}}$. 
Note that, in contrast with these works, we have considered here very small temperatures, $T\ll T_V$. To resort to a 
non-local algorithm to accelerate the convergence of simulations is therefore essential.
Note also that at the intermediate temperatures where the necessity of using global Monte Carlo schemes is less important
we have also found a clear improvement associated with the use of cluster algorithms.

The main result of this paper is that we have found some striking differences in the 
behavior of the spin-stiffnesses as a function of the linear size for the two models,
with and without topological excitations. 
At very low temperatures the temperature-rescaled spin-stiffness, $\widetilde\rho={\rho/ T}$ (the natural 
quantity to consider, see below) of both models displays the characteristic behavior:
\begin{equation}
\widetilde\rho={\rho\over T} \sim {1\over 4\pi}\ln {\xi \over L}
\label{formula1}
\end{equation} predicted by the perturbative RG approach of
the NL$\sigma$ model$^{\cite{aza7}}$. This is, of course, expected since the  spin-wave contents of  both 
models  are identical. We call the regime corresponding to this range of temperatures the ``spin-wave'' regime. 
Note that, for one value of the temperature, this asymptotic scaling of $\widetilde\rho$ with 
respect to $\ln {L}$ has already been confirmed directly on the HAFT model by Southern 
and Young$^{\cite{southern}}$. At higher, but still low, temperatures the two models 
begin to display different behaviors. 
While the spin-stiffness of the topologically non-trivial model still displays 
the previous characteristic behavior, its absolute magnitude with respect to the trivial model is found 
to decrease quite rapidly as a function of the temperature. 
We propose to refer to this regime as an ``almost-spin-wave'' regime, a regime where the only significant 
effect of vortices is just to shift down the value of $\rho/T$.
Next, at higher temperatures we enter a regime called here the ``vortex'' regime
where the vortices play a major role.  In this regime, the spin-stiffness looses its regular behavior. It exhibits
large fluctuations around its mean value with the presence
of ``plateaux'' and abrupt jumps as a function of the linear size. 
Nevertheless, by considering the global behavior of the curve it is 
still possible to define 
some effective linear regime as a function of $\ln L$ similar to that described by Eq.(\ref{formula1}). However, 
in contrast with the spin-wave regime, the slope of the spin-stiffness
is no longer constant ($1/4\pi)$ and is found to increase quite rapidly as a function of the temperature.
It is remarkable that this 
regime is  observable only within a narrow range of temperatures. 
At slightly higher temperatures, the curve of the spin-stiffness recovers 
a much more conventional behavior: smooth decrease as a 
function of the size and cancellation of $\rho$ 
at some finite lattice size corresponding to some finite correlation length.
Regarding the theoretical interpretation of our results, we show that the very low-temperature regime is 
in full agreement with the RG predictions. The so-called ``almost-spin-wave'' 
and ``vortex'' regimes are much more puzzling.
However, it is shown that the most salient features induced by the topological defects in these regimes can be 
rather well reproduced using some simple phenomenological RG equations which combine 
the topology of the XY model and the spin-wave content of the $O(4)$ model. 

The organization of the paper is as follows. In Section II, the various actions of the lattice models are 
presented. In Section III, the expressions of the spin-stiffnesses suitable for Monte Carlo 
simulations are given. 
In Section IV, we present briefly the Wolff-Swendsen-Wang algorithm used. 
Our results are given in Section V. 
In this latter section we present the behavior of 
the spin-stiffnesses as a function of the lattice size in the 
various temperature regimes going from low- to high-temperatures. Finally, we present 
in the last section our first attempt toward a theoretical interpretation of the effect of 
the vortices in the almost-spin-wave and vortex regimes.

\section{The lattice models}

\subsection{The $SO(3)\otimes O(2)$ top model}

Our first step is to map the HAFT model into an equivalent non-frustrated  one.
As shown by Dombre and Read$^{\cite{dombre1}}$ 
and Azaria {\it et al.}$^{\cite{aza4}}$ the long-distance effective hamiltonian 
of the HAFT model consists in a system of classical
interacting tops. This can be understood from the fact that the 
$120^{\circ}$ structure of the spins of the HAFT model
in the ground state fully breaks the $SO(3)$ symmetry so that the order parameter 
is a rotation matrix $R\in SO(3)$, a classical top. As in the non-frustrated case, once the theory is reformulated  
in terms of the order parameter, the
effective interaction becomes ferromagnetic.  
The hamiltonian of the top model thus reads$^{\cite{dombre1,aza4}}$:
\begin{equation}
H_1=-\sum_{<i,j>}\ \hbox {Tr}\left(P R_i^{-1} R_j \right)
\label{action1}
\end{equation}
where $R_i$ is a rotation matrix of $SO(3)$ defined on site $i$ and $P$=diag$(p_1,p_1,p_3)$ is a diagonal
 matrix of positive coupling constants which represents the interaction strengths between the different axes of the tops.
 Note that the temperature has been included in the $p_i$'s. The HAFT model corresponds to the special case $p_3=0$$^{\cite{dombre1,aza4}}$.  
The $SO(3)$ symmetry of the HAFT is realized here through  the rotational invariance of  hamiltonian ($\ref{action1}$)
  under left global $SO(3)$ rotations $R_i\to UR_i$, $U\in SO(3)$.  With the matrix $P$ considered
 here it is also invariant under the $O(2)$ group of right global transformations: $R_i\to R_iV$ that commute with 
the matrix $P$.  Thus, hamiltonian ($\ref{action1}$)  is invariant under the group $G=SO(3)\otimes O(2)$. 
This left $O(2)$ group is reminiscent of the $C_{3v}$ symmetry of the triangular lattice. Note that it will be convenient
 in the following to consider the case $p_3\ne 0$ since the hamiltonian made with this $P$ is the general one invariant 
under $G=SO(3)\otimes  O(2)$ and that, as well known$^{\cite{aza4,aza7}}$, the condition $p_3=0$ is not preserved by renormalization.  

The symmetry breaking pattern described by hamiltonian (\ref{action1}) is   $G=SO(3)\otimes O(2)$ broken down 
to $H = O(2)$: 
\begin{equation}
{G\over H}={SO(3)\otimes O(2)\over O(2)}\equiv SO(3)
\end{equation}
where the notation $\equiv$ means that $G/H$ is topologically isomorphic to $SO(3)$. 
The symmetry breaking pattern thus corresponds to a fully broken $SO(3)$ group. For the original HAFT model,
 this symmetry breaking pattern is $G/H=SO(3)\otimes C_{3v}/C_{3v}$ and is thus identical to that given
 by Eq.(\ref{action1}).  This is the reason why the substitution of the discrete $C_{3v}$ by the continuous $O(2)$
 one in hamiltonian (\ref{action1}) is harmless. It is interesting to note the identity: 
$SO(3)=SO(4)/(SO(3)\otimes Z_2)=S_3/Z_2$, $S_3$ being the 3-sphere, since it shows that the model of
 tops (\ref{action1}) is equivalent to that of four-component spins living on the four-dimensional unit sphere 
with antipodal points identified. This will allow us, in the following, to build a vector model equivalent to the
 preceding matrix one and suitable for Monte Carlo simulations. Note finally that when $p_1=p_3$
 the symmetry group is  enlarged to  $G=SO(3)\otimes SO(3)$ and the symmetry breaking pattern is 
 $G/H=SO(3)\otimes SO(3)/SO(3)$. 

\subsection{The $SU(2)\otimes U(1)$ top model}

We now build the topologically trivial counterpart of the previous top model. 
We want to preserve the spin-wave part of the model while discarding the topological excitations. 
Since the spin-wave excitations correspond to small fluctuations
of the order parameter, they only probe the local structure of the order parameter space $G/H$ 
and not its global -- topological -- structure. This local structure is itself completely
determined by the Lie algebras of $G$ and $H$$^{\cite{aza4,friedan}}$. We thus need a model defined by an order parameter
space $G'/H'$ locally isomorphic to $G/H$ and topologically trivial. This is obtained by considering the covering group $SU(2)$ 
of $SO(3)$. The relevant model is thus built on the manifold $SU(2)\otimes U(1)/U(1)$.
The most general hamiltonian invariant under $SU(2)\otimes U(1)$ writes:
\begin{equation}
H_2=-\sum_{<i,j>}\left\{ 2 (p_1+p_3)\ 
\hbox {Tr}\ g_i^{-1} g_j + {1\over 2}(p_1-p_3)\ \left(\hbox{Tr}\ \sigma_3\ g_i^{-1} g_j\right)^2\right\}
\label{action2}
\end{equation}
where $g_i \in SU(2)$ and $\sigma_3$ is the third Pauli matrix. The first term in this hamiltonian is clearly invariant under the  cross product of
 a left $SU(2)$ group and  a right $SU(2)$ group: $g_i\to M g_i N$, $M,N\in SU(2)$. The second 
one explicitly breaks the right $SU(2)$ down to a right $U(1)$ so that the hamiltonian is
 generically $SU(2)\otimes U(1)$ invariant. This $U(1)$ symmetry corresponds to the $O(2)$ 
symmetry of hamiltonian ($\ref{action1}$).  

The symmetry breaking pattern described by hamiltonian (\ref{action2}) is:
$G'=SU(2)\otimes U(1)$ broken down to $H' = U(1)$:
\begin{equation}
{G'\over H'}={SU(2)\otimes U(1)\over U(1)}\equiv SU(2)
\end{equation}
so that it corresponds to a fully broken $SU(2)$ group. 
Again, it is interesting for the following to note the identity $SU(2)=SO(4)/SO(3)=S_3$
 which means that the model (\ref{action2}) is equivalent to that of four-component spins living on the
 four-dimensional unit sphere.  Again, when $p_1=p_3$, the symmetry group is enlarged
 to $G'=SU(2)\otimes SU(2)$ and the symmetry breaking pattern becomes  ${SU(2)\otimes
 SU(2)/SU(2)}$. 
Note finally  that the choice of coupling constants in ($\ref{action2}$) is such that the two models ($\ref{action1}$)  and ($\ref{action2}$) have the same temperature scale.

The models corresponding to ($\ref{action1}$) and ($\ref{action2}$) have, by construction, the same spin-wave dynamics but can nevertheless strongly differ when excitations associated with the topology are activated.  The $SU(2)\otimes U(1)$ model being topologically trivial, i.e. $\pi_1(SU(2))=0$, we expect that a Monte Carlo study 
of this model will be well reproduced
by a pure spin-wave approach.  We show in the following that this is 
indeed what happens: as in the topologically trivial $O(N)/O(N-1)$ ferromagnetic spin systems, 
the critical properties of the $SU(2)\otimes U(1)$ model are in perfect 
agreement with the perturbative RG
predictions made on the continuous limit of the top model, a NL$\sigma$ model. On the other hand, the $SO(3)\otimes O(2)$ model being 
topologically non-trivial, some disagreement between the 
perturbative and Monte Carlo approaches at sufficiently high temperatures are found, as expected. 

\subsection{The vectorial version of the $SO(3)\otimes O(2)$ model}

As already mentioned in the introduction, the cluster algorithms are easier to implement 
for spins than for matrices. 
We thus need vectorial versions of our hamiltonians. 
This is achieved by using  the decomposition of a rotation matrix $R_i$ of $SO(3)$ in terms of a four-component
 unit vector $\vec{S_i}=(S^0_i, {\pmb{$S$}}_i)=(S^0_i, S^1_i, S^2_i, S^3_i)$:
\begin{equation}
R^{kl}_i=2\left(S^k_iS^l_i-{1\over 4}
\delta_{kl}\right)+2\epsilon_{klm}S^0_iS^m_i+2\left({S^0_i}^2-{1\over 4}\right)\delta_{kl}\ \ .
\label{decomposition}
\end{equation}
The hamiltonian ($\ref{action1}$) then takes the form 
\begin{equation}
\begin{array}{l}
{H_1'}=-{\displaystyle \sum_{<i,j>}}\left\{ {{4p_1}\left(\left(\vec{S}_{i}. \vec{S}_{j}\right)^2-{\displaystyle {1\over 4}}\right)}\right.
\\
\\
\hskip1cm+\ {4\left(p_3-p_1\right)\bigg[\left(S_{i}^0 S_{j}^0+S_{i}^3 S_{j}^3\right) \left(S_{i}^1 S_{j}^1+S_{i}^2 S_{j}^2\right)+\left(S_{i}^0 S_{j}^3-S_{j}^0 S_{i}^3\right) \left(S_{i}^1 S_{j}^2-S_{i}^2 S_{j}^1\right)\bigg.}\\
\\
\left. \hskip1cm+{ \left.{\displaystyle{1\over 4}}\left({S_{i}^0}^2+{S_{i}^3}^2-{S_{i}^1}^2-{S_{i}^2}^2\right)
\left({S_{j}^0}^2+{S_{j}^3}^2-{S_{j}^1}^2-{S_{j}^2}^2\right)\right]} \right\}
\end{array}
\label{action3}
\end{equation}
The first term of (\ref{action3}) represents the hamiltonian of a system of spherical tops -- i.e. $SO(3)\otimes SO(3)\simeq SO(4)$ symmetric --  for which $p_1=p_3$. 
This term is also invariant under a {\it local} -- gauge -- $Z_2$ group. This $Z_2$ 
symmetry expresses the non trivial topological character of the $SO(3)$ group. The  first term of (\ref{action3})
 is also known as the hamiltonian of the $RP^3=SO(4)/(SO(3)\otimes Z_2)$ model which expresses the
 isomorphism between the manifolds $SO(3)$ and $RP^3$. This model and, more generally, the $RP^N$ models
 for general $N$ have been extensively studied and the  question of the nature of their continuum limit
$^{\cite{zumbach8,caracciolo,hasenbusch1,hasenbusch2,niedermayer,catterall}}$ strongly debated, 
also in connection with topological defects. The
  second term of hamiltonian (\ref{action3}) also displays the $Z_2$ local symmetry but breaks 
the global $SO(4)$  symmetry so that the  hamiltonian is generically globally $SO(3)\otimes O(2)$ 
and locally $Z_2$ symmetric. 

\subsection{The vectorial version of the $SU(2)\otimes U(1)$ spin  model}

It is also possible to express hamiltonian  (\ref{action2}) in terms of four-component vectors $\vec{S}_i$.
Using the decomposition of a $SU(2)$ matrix:
\begin{equation}
g_i=S_{i}^0 +i {\pmb{$\sigma$}}.{\pmb{$S$}}_i
\end{equation}
$\sigma_k$, $k=1, 2, 3$ being  the Pauli matrices, hamiltonian (\ref{action2}) writes:
\begin{equation}
H_2'=-\sum_{<i,j>}4(p_1+p_3)\ 
 \vec{S}_{i}. \vec{S}_{j}+2(p_1-p_3)\left(S_{i}^0 S_{j}^3-S_{j}^0 S_{i}^3-S_{i}^2 
S_{j}^1+S_{j}^2 S_{i}^1\right)^2\ .
\label{action5}
\end{equation}
In this expression the first term is $O(4)$ globally invariant and
corresponds to the hamiltonian of a four-component ferromagnet  whereas the second term breaks
this symmetry. The hamiltonian (\ref{action5}) is thus generically globally $SU(2)\otimes U(1)$ symmetric. 
Note that the $Z_2$ local symmetry has now disappeared. This is a consequence of the trivial
topological character of $SU(2)$.
Note also that the scale of temperature has been chosen so that both Hamiltonians $H_1'$, Eq.(\ref{action3}),
and $H_2'$, Eq.(\ref{action5}), have the same linearized spin-wave form in the symmetric case ($p_1=p_3$), namely
$H_{1/2}'= -\sum_{<i,j>}8p_1 \delta{\vec S}_{i}.\delta{\vec S}_{j}$ 
where $\delta{\vec S}$ represents the spin deviation from the reference vector.
Finally, remark that the Jacobians resulting from the change of variables: 
matrices $\rightarrow$ spins in both $SO(3)\otimes O(2)$ 
and $SU(2)\otimes U(1)$ are trivial and thus do not contribute to the free energy.

\section{The Spin-stiffnesses}

\subsection{The spin-stiffnesses of the lattice models}

The spin-stiffness $\rho_{\alpha}$ to be computed numerically is defined as the free energy increment 
under twisting the boundary conditions, for instance in the $x$ direction around the direction $\alpha$. 
This is realized by coupling the system with two walls of tops: $R(x=0)=R_1$ and $R(x=L)=R_2$, $R_2$ being
deduced from $R_1$ by a rotation of angle $\theta_{\alpha}$ around the direction $\alpha$ and 
by measuring the variation of the free energy with respect to $\theta_{\alpha}$:
\begin{equation}
\rho_{\alpha}={\partial^2 F(\theta_{\alpha})\over \partial\theta_{\alpha}^2}
 {\bigg{\vert}_{\theta_{\alpha}=0}}\ \ .
\label{stiff}
\end{equation}
For a system with partition function:
\begin{equation}
Z=\displaystyle \sum_{\left[R_i\right]}\ e^{-\displaystyle H}
\label{partition}
\end{equation}
we have:
\begin{equation}
\rho_{\alpha}=-T\left[-\bigg\langle{\partial H\over \partial \theta_{\alpha}}\bigg\rangle^2
-\bigg\langle{\partial^2 H\over \partial \theta_{\alpha}^2}\bigg\rangle+\bigg\langle\left({\partial H\over \partial \theta_{\alpha}}\right)^2\bigg\rangle\right]_{\theta_{\alpha}=0}\ .
\label{stiffi}
\end{equation}

Since $H$ is even in the $\theta_{\alpha}$'s, the average value of ${\partial H/ \partial \theta_{\alpha}}$ 
is equal to zero and only the two last terms of Eq.(\ref{stiffi}) need to be computed.

\subsubsection{The spin-stiffnesses of the $SO(3)\otimes O(2)$ model.}

In principle we have to compute the different average values in (\ref{stiffi}) from the partition function:
\begin{equation}
Z=\displaystyle\sum_{\left[R_i\right]}\ \displaystyle 
\exp{(\displaystyle\sum_{<i,j>} \hbox{Tr}P R_i^{-1}R_j  )}
\label{partitionbis}
\end{equation}
constrained by the boundary conditions:
\begin{equation}
\left\{
\begin{array}{lll}
\displaystyle{R(x=0)=R_0}
&&\\
\displaystyle{R(x=L)=R_0 \ e^{\displaystyle i\theta_{\alpha} T_{\alpha}}}\ 
\end{array}
\right.
\label{boundary}
\end{equation}
where $R_0$ is a rotation matrix of reference (e.g. $R_0=\BBU$), $T_{\alpha}$ is the generator of rotation 
around the $\alpha$ direction, and $\theta_{\alpha}$ the angle of rotation 
(Eq.(\ref{boundary}) must be understood {\it without} the sum over $\alpha$).
However, in practice, the presence of derivatives with respect to $\theta_\alpha$ in Eq.(\ref{stiffi}) 
as well as the fact that the cluster algorithm is implemented with
spins, makes expression (\ref{stiffi}) not suitable for our simulations. We proceed in two steps to reformulate
the model in a numerically convenient way. First, to get rid of the derivatives, we compute them analytically and rewrite the average values in (\ref{stiffi}) as $\theta_\alpha$-independent quantities. 
To do this we  decompose $R_i$ into a zero temperature part -- $R_i^{cl}$ -- and a fluctuation part -- $h_i$ -- :
\begin{equation}
R_i=R_i^{cl}\ h_i
\label{rot}
\end{equation}
where both $R_i^{cl}$ and $h_i$ belong to $SO(3)$.
In Eq.$(\ref{rot})$, $R_i^{cl}$ is by definition a solution of the classical equations of motion and thus reads:
\begin{equation}
R_i^{cl}=e^{-\displaystyle i\ \theta_\alpha T_\alpha {x_i\over L}}
\label{rotation2}
\end{equation}
and the $h_i$'s satisfy the boundary
 conditions: 
\begin{equation}
h(x=0)=h(x=L)=\BBU. 
\label{boundaryh}
\end{equation}
We thus have:
\begin{equation}  
\begin{array}{lll}
\displaystyle{{\partial H\over \partial \theta_\alpha}{\bigg{\vert}_{\theta_\alpha=0}}=
{1\over L}\sum_{<i,j>,k,l,m} p_k\  \epsilon_{\alpha l m }h_i^{kl} h_j^{km} (x_i - x_j)}\ .
&&\\
\\
\displaystyle{{\partial^2 H\over \partial \theta_\alpha^2}{\bigg{\vert}_{\theta_\alpha=0}}=
{1\over L^2}\sum_{<i,j>,k,l,m} p_k 
[\delta_{lm}h_i^{kl} h_j^{km}-h_i^{k\alpha}h_j^{k\alpha}]
 (x_i - x_j)^2}\ .
\end{array}
\label{moumoudel}
\end{equation}
The average values in Eq.(\ref{stiffi}) must now be computed with:
\begin{equation}
Z=\displaystyle \displaystyle\sum_{\left[h_i\right]}\ 
\displaystyle \exp{\bigg( \displaystyle\sum_{<i,j>} 
\hbox{Tr} P h_i^{-1}  h_j \bigg)}
\label{partition2}
\end{equation}
with the boundary conditions $(\ref{boundaryh})$. At this stage, everything is written in terms of the $h$'s. Thus we can now perform the second step of our derivation that consists in using in Eq.(\ref{boundaryh}), (\ref{moumoudel}) and (\ref{partition2}) the same decomposition as in Eq.(\ref{decomposition}) but now for the $h_i$'s:
 \begin{equation}
{
h_i^{kl}=2\left(S_i^kS_i^l-{1\over 4}\delta_{kl}\right)
+2\epsilon_{klm}S_i^0S_i^m+2\left({S_i^0}^2-{1\over 4}\right)\delta_{kl}\ }\ .
\label{decomposition2}
\end{equation}
Since now all thermal average values are entirely expressed in terms of four-component spins, the cluster algorithm can be implemented to compute the spin-stiffnesses.

\subsubsection{The spin-stiffnesses of the $SU(2)\otimes U(1)$ model.}

The same method can be employed for the $SU(2)\otimes U(1)$ model for which we have:
\begin{equation}
Z=\displaystyle\sum_{\left[g_i\right]}\ \displaystyle 
\exp{ \bigg( \displaystyle\sum_{<i,j>} 2 (p_1+p_3)\ \hbox {Tr}\ g_i^{-1} g_j + 
{1\over 2}(p_1-p_3)\ \left(\hbox{Tr}\ \sigma_3\ g_i^{-1} g_j\right)^2 \bigg)}
\label{partitionter}
\end{equation}
with, again, the fixed boundary conditions:
\begin{equation}
\left\{
\begin{array}{lll}
\displaystyle{g(x=0)=g_0}
&&\\
\displaystyle{g(x=L)=g_0 \ e^{\displaystyle i{\theta_\alpha} {\sigma_{\alpha}\over 2}}}
\end{array}
\right.
\label{boundary2}
\end{equation}
where $g_0$ is a rotation matrix of reference of $SU(2)$ (e.g. $g_0=\BBU$), $\sigma_{\alpha}$ is a Pauli matrix, 
and $\theta_\alpha$ the angle of rotation (Again in Eq.(\ref{boundary2}) there is no sum over $\alpha$). 

As in the $SO(3)$ case, we make the decomposition in classical and fluctuating parts:
\begin{equation}
g_i=g_i^{cl}\ h_i
\label{rot2}
\end{equation}
with:
\begin{equation}
g_i^{cl}=e^{\displaystyle i\ \theta_\alpha {\sigma_\alpha\over 2} {x_i\over L}}
\label{rotation3}
\end{equation}
and $h$ satisfying: 
\begin{equation}
h(x=0)=h(x=L)=\BBU\ .
\label{boundaryh2}
\end{equation}

The different terms of ($\ref{stiffi}$) are separated
 into $SO(4)$ and $SU(2)\otimes U(1)$ symmetric parts. 
Writing  $S=S_{SO(4)}+S_{SU(2)\otimes U(1)}$ and using the decomposition:
\begin{equation}
h_i=S_{i}^0 +i {\pmb{$\sigma$}}.{\pmb{$S$}}_i\ ,
\end{equation}
we have with obvious notations:

\begin{equation}
\begin{array}{lll}
\displaystyle{{\partial H_{O(4)}\over \partial {\scriptsize{\left(\begin{array}{c}
\theta_1\\
\theta_2\\
\theta_3\\
\end{array}\right)}}}
{\Bigg{\vert}_{\theta_{\alpha}=0}}= {2\over L} (p_1+p_3)\sum_{<i,j>}
\left(
\begin{array}{c}
S_i^0S_j^1-S_i^1S_j^0+S_i^3S_j^2-S_i^2S_j^3\\
S_i^0S_j^2-S_i^2S_j^0+S_i^1S_j^3-S_i^3S_j^1\\
S_i^0S_j^3-S_i^3S_j^0+S_i^2S_j^1-S_i^1S_j^2\\
\end{array}
\right)
(x_j-x_i)}
&&\\
\\
\displaystyle{{\partial H_{SU(2)\otimes U(1)}\over \partial\  { \scriptsize{\left(\begin{array}{c}
\theta_1\\
\theta_2\\
\theta_3\\
\end{array}\right)}}}{\Bigg{\vert}_{\theta_{\alpha}=0}}={2\over L}(p_3-p_1)  \sum_{<i,j>} \omega^3_{ij}
\left(
{\begin{array}{c}
S_i^3S_j^1+S_i^1S_j^3-S_i^2S_j^0-S_i^0S_j^1\\
S_i^3S_j^2+S_i^2S_j^3+S_i^1S_j^0+S_i^0S_j^2\\
S_i^0S_j^0+S_i^3S_j^3-S_i^1S_j^1-S_i^2S_j^2\\
\end{array}}
\right)
(x_j-x_i)}       
&&\\
\\
\displaystyle{{\partial^2 H_{O(4)}\over \partial\  {\scriptsize{\left(\begin{array}{c}
\theta^2_1\\
\theta^2_2\\
\theta^2_3\\
\end{array}\right)}}}{\Bigg{\vert}_{\theta_{\alpha}=0}}}= {1\over L^2}(p_1+p_3)  \sum_{<i,j>}
\left(
\begin{array}{c}
\vec{S}_i . \vec{S}_j\\
\vec{S}_i . \vec{S}_j\\
\vec{S}_i . \vec{S}_j\\
\end{array}
\right)
(x_j-x_i)^2
&&\\
\\
\displaystyle{{\partial^2 H_{SU(2)\otimes U(1)}\over \partial\  {\scriptsize{\left(\begin{array}{c}
\theta^2_1\\
\theta^2_2\\
\theta^2_3\\
\end{array}\right)}}}{\Bigg{\vert}_{\theta_{\alpha}=0}}}=
{1\over L^2}(p_1-p_3)  \sum_{<i,j>} 
\left(
\begin{array}{c}
{(S_i^3S_j^1+S_i^1S_j^3-S_i^2S_j^0-S_i^0S_j^2)}^2 - { \omega^3_{ij}}^2 \\
{(S_i^3S_j^2+S_i^2S_j^3+S_i^1S_j^0+S_i^0S_j^1)}^2 - { \omega^3_{ij}}^2 \\
{(S_i^0S_j^0+S_i^3S_j^3-S_i^1S_j^1-S_i^2S_j^2)}^2 - { \omega^3_{ij}}^2 \\
\end{array}
\right)
(x_j-x_i)^2
\end{array}
\label{moumoudel2}
\end{equation}
with:
\begin{equation}
{
\omega^3_{ij}= -S_i^0S_j^3+S_i^3S_j^0-S_i^1S_j^2+S_i^2S_j^1\ .
}
\end{equation}
The spin-stiffnesses can now be computed with:
\begin{equation}
Z=\displaystyle \sum_{\left[h_i\right]}\ \displaystyle 
\exp{\bigg(\displaystyle\sum_{<i,j>}2(p_1+p_3)\ \hbox {Tr}\ h_i^{-1} h_j + {1\over 2}(p_1-p_3)\ \left(\hbox{Tr}\ 
\sigma_3\ h_i^{-1} h_j\right)^2\bigg)}\ 
\label{partition3}
\end{equation}
with the set of equations (\ref{moumoudel2}) 
and the boundary conditions (\ref{boundaryh2}) or, in terms of the four-component vector $\vec{S}$:
\begin{equation}
\vec{S}(x=0)=\vec{S}(x=L)=(1,0,0,0)\ .
\end{equation}

\subsection{The spin-stiffnesses of the NL$\sigma$ models}

The spin-stiffnesses can be analytically computed from the continuum versions of the top models, which is a Non Linear Sigma (NL$\sigma$) model. They identify with the effective coupling constants  of 
the  NL$\sigma$ model at scale $L$, renormalized by thermal fluctuations.  Let us recall that the perturbative
 treatment of the NL$\sigma$ model takes only into account the spin-wave part of the spin-stiffnesses. 
Indeed, the $\beta$ functions of any NL$\sigma$ model are completely determined  by the
 local properties, i.e. by the metric, of the manifold $G/H$ while they are
 insensitive to its global, i.e. topological, structure. Since the metric  itself
 is completely determined by the Lie Algebras of $G$ and $H$, the perturbative 
$\beta$ functions -- and thus the behavior of the spin-wave part of the spin-stiffnesses -- 
of the $SO(3)\otimes O(2)$ and  $SU(2)\otimes U(1)$   
 NL$\sigma$ models are identical by construction. They are given at two-loop order by:$^{\cite{aza4,aza7}}$
\begin{equation}
\left\{
\begin{array}{lll}
{\displaystyle{
{\partial\widetilde{\rho}_1(l)\over\partial l}}}& \hspace{-0.2cm}={\displaystyle{-{1\over 2\pi}+
\frac{1}{4\pi}
\frac{\widetilde{\rho}_3(l)}{\widetilde{\rho}_1(l)} -
\frac{5}{32\pi^2}\frac{{\widetilde{\rho}_3(l)}^2}{{\widetilde{\rho}_1(l)}^3}+\frac{3}{8\pi^2}\frac{{\widetilde{\rho}_3(l)}}
{{\widetilde{\rho}_1(l)}^2}}}-{\displaystyle{\frac{1}{4\pi^2\widetilde{\rho}_1(l)}}}\\
\\
{\displaystyle{\frac{\partial\widetilde{\rho}_3(l)}{\partial l}}} & \hspace{-0.2cm}=
-{\displaystyle{
\frac{1}{4\pi}\frac{{\widetilde{\rho}_3(l)}^2}{{\widetilde{\rho}_1(l)}^2} -\frac{1}{32\pi^2}
\frac{{\widetilde{\rho}_3(l)}^3}{{\widetilde{\rho}_1(l)}^4}}}\ 
\end{array}
\right.
\label{cl433}
\end{equation}
with $l=\displaystyle\ln{L/a}$  where $L$ is the system size, $a$ the lattice spacing, the initial 
conditions of the RG flow being given by:
\begin{equation}
\left\{
\begin{array}{lll}
\displaystyle{\widetilde{\rho}_1(l=0)=p_1+p_3}
\\
\displaystyle{\widetilde{\rho}_3(l=0)=2p_1\ .}
\end{array}
\right.
\label{stiti}
\end{equation}
Note that the RG equations (\ref{cl433}) are written in terms of quantities that contain, 
in their definition, the temperature, as it is clear from Eq.(\ref{stiti}) since the temperature 
is included in the $p_i$'s. 
The Monte Carlo counterpart of the $\widetilde{\rho}_{\alpha}$'s are thus given by 
the $\rho_{\alpha}/T$'s previously defined.

In contrast with the $O(N)$ case, there are {\it a priori} 
three different spin-stiffnesses in our case but, in fact,  
only two are  independent since the left $O(2)$ symmetry constrains two of them to be
 identical. Note that, asymptotically, i.e. for $L\gg  a$, the behavior of the $\widetilde\rho_{\alpha}$'s 
is given by the infrared limit of the flow equations. It is easy to show that, 
in this limit $\widetilde{\rho}_1\to \widetilde{\rho}_3$ and  the model becomes
 effectively $SO(3)\otimes SO(3)\sim SO(4)$ symmetric. This is the known phenomenon 
of enlarged symmetry$^{\cite{aza1,aza4}}$. We recover therefore 
the universal scaling of the spin-stiffness of a  ferromagnetic $N=4$ vector
 model. At leading order we have:$^{\cite{aza7}}$ 
\begin{equation}
\widetilde{\rho}_1\sim \widetilde{\rho}_3\sim {1\over 4\pi}\ln {\xi \over L}\ 
\label{stiffcorrelation}
\end{equation}
where $\xi$ is the correlation length.
Note, of course, that since the spin-stiffnesses in Eq.(\ref{cl433})
 are calculated perturbatively, they can only be valid in a limited range of {\it low}
 temperatures where the perturbation theory at two-loop order is meaningful.

From the preceding analysis, it should be clear that for a topologically trivial model, 
the spin-wave part of the spin-stiffness $\widetilde{\rho}_{\alpha}$  
identifies with  $\rho_{\alpha}/T$, so that we can expect that this last quantity  follows the RG equations calculated by perturbation theory.  Indeed, for the $O(N)$ model,
 it has been checked$^{\cite{aza9}}$  that, at sufficiently low temperature, the spin-stiffness $\rho_{\alpha}/T$ follows the RG equations calculated by means of the $O(N)$ NL$\sigma$ model up to two loop order$^{\cite{brezin8,chakravarty3}}$.
 In the same way, we expect the different spin-stiffnesses $\rho_{\alpha}/T$ of the
 lattice $SU(2)\otimes U(1)$ model calculated by Monte Carlo simulation to follow
 the RG equations (\ref{cl433}) in a large range of low temperatures since it is topologically trivial.
 On the other hand, we expect the behavior of the spin-stiffnesses $\rho_{\alpha}/T$ 
of the lattice $SO(3)\otimes O(2)$ model to agree with (\ref{cl433}) at very low
 temperature, i.e. below $T_V$, where the topological defects are not activated, but
 to disagree with the perturbative RG predictions near and above the cross-over temperature $T_V$. 

\section{Wolff-Swendsen-Wang Algorithm}

The simulations presented in this paper are based on a generalization of
the Wolff-Swendsen-Wang \cite{swendsen},\cite{wolff},\cite{wolff2},
\cite{wolff3} algorithm to $N$-vector models as presented by 
Caracciolo {\it et al.} in \cite{caracciolo2}. The method is based on an 
embedding of Ising spins ${\epsilon}$ into the $N$-component 
(here $N=4$) continuous spins {$\vec{S}$ according to:
\begin{equation}
\vec{S}_i= \vec{S}_i^{\perp} + \epsilon_i |\vec{S}_i^{\parallel}| \vec{r}
\label{embed}
\end{equation}
where $\vec{r}$ is a unit vector chosen randomly on the sphere $S^3$, 
$\vec{S}_i^{\perp}= \vec{S}_i - (\vec{S}_i \cdot \vec{r}) \vec{r} $ and 
$\vec{S}_i^{\parallel}=(\vec{S}_i \cdot \vec{r}) \vec{r}$ are the components 
of the spin vector perpendicular and parallel to the unit vector $\vec{r}$,
the Ising variable $\epsilon_i$ being given by $\epsilon_i = 
sgn (\vec{S}_i \cdot \vec{r}) =\pm 1$. Once this embedding is done, our 
initial hamiltonian written in terms of continuous spins 
-- here, hamiltonians (\ref{action3}) and (\ref{action5}) -- can be rewritten as 
a generalized random-bond Ising model. In the  Monte Carlo simulation 
the spin variables of this new problem are updated using an efficient 
non-local algorithm for Ising variables (e.g. the standard Swendsen-Wang 
algorithm). To flip the Ising variable $\epsilon_i$ corresponds to make 
a reflection of the vector $\vec{S}_i$ in the hyperplane perpendicular
 to $\vec{r}$. A necessary condition to get an efficient Wolff-type 
algorithm is 
that this transformation  preserves the total energy of the system. In that 
case its application to a large set of spins costs 
only a surface energy and large-scale changes in the spin configuration 
are possible. Here, such a condition is verified only for the symmetric part
($p_1=p_3$) of the hamiltonians. As a consequence, we have chosen to perform 
our simulations with the reference symmetric hamiltonians.
More precisely, for the $SO(3)\otimes O(2)$ model we consider:
\begin{equation}
H{^{(0)}}=-{\displaystyle \sum_{<i,j>}}{{4p_1} \bigg( \left(\vec{S}_{i}. 
\vec{S}_{j}\right)^2-{\displaystyle {1\over 4}} \bigg)}\\
\label{hamilt1}
\end{equation}
and for the  $SU(2)\otimes U(1)$ model we take
\begin{equation}
H{^{(0)}}=-{\displaystyle \sum_{<i,j>} } 8p_1  \vec{S}_{i}.\vec{S}_{j}\\
\label{hamilt2}
\end{equation}
Note that both hamiltonians reduce to the same hamiltonian in the spin-wave approximation 
($H_{SW}= -\sum_{<i,j>}8p_1 \delta{\vec S}_{i}.\delta{\vec S}_{j}$ where $\delta{\vec S}$ represents
the spin deviation from the reference vector).
Note also that the zero-temperature ground-state energies are different for the two hamiltonians. 
This is not important since the various quantities computed in this work do not 
depend on this reference energy.

Calculation of exact properties associated with the full hamiltonians 
are done by reweighting appropriately the Monte Carlo averages. 
Let us write $\bar{Q}$ the average of an arbitrary function $Q(S)$
of the spin configuration ${S}$
\begin{equation}
{\bar Q} = \int d\vec{S}_1 \ldots \int d\vec{S}_M Q(S) 
e^{-H(S)/T} / Z
\label{averageQ}
\end{equation}
where $M$ is the total number of spins considered, $H(S)$ is the 
exact hamiltonian, and $Z$ the partition function. We re-express
${\bar Q}$ as follows
\begin{equation}
{\bar Q} = \int d\vec{S}_1 \ldots \int d\vec{S}_M Q(S) 
e^{-V(S)/T} e^{-H^{(0)}(S)/T} / 
\int d\vec{S}_1 \ldots \int d\vec{S}_M
e^{-V(S)/T} e^{-H^{(0)}(S)/T} 
\label{averageQbis}
\end{equation}
where $V$ is the difference between the exact and reference hamiltonian,
$V\equiv H-H^{(0)}$
The Monte Carlo averages are performed over the set of spin 
configurations distributed according to the Boltzmann weight  
associated with the reference hamiltonian (denoted here as $\langle\langle\cdots{\rangle\rangle}_{0}$)
\begin{equation}
{\bar Q} = \langle\langle Q(S) e^{-V(S)/T} {\rangle\rangle}_0 /
\langle\langle e^{-V(S)/T} {\rangle\rangle}_0.
\label{average}
\end{equation}
By using such a procedure we are sure that at the symmetric 
point ($p_1=p_3$) our 
simulations are free from the critical slowing down problem 
(see, Ref.(\cite{caracciolo2})). Away from the symmetric point, the 
situation is less clear. There is a subbtle interplay between the 
loss due to the undesirable fluctuations
of the weight in averages (a growing source as the asymmetric parameter 
$\Delta p \equiv p_1-p_3$ is increased) and the gain obtained from 
the treatment of the large-scale 
collective-mode moves issued from the reference hamiltonian. In 
practice, we have found that realistic calculations can be done only for a
small value of $\Delta p$ with a maximum value of about 0.2. 
Although we have not made a systematic study of the effective (associated 
with $H$ and not $H^{(0)}$) dynamical exponent, we are quite confident 
that, in the regime $\Delta p < 0.2$,
the convergence of the estimators at the very low temperatures 
we have considered is very satisfactory. 

In order to compute the spin-stiffnesses we have used formulas 
(\ref{stiffi}), (\ref{moumoudel})  and (\ref{moumoudel2}) presented in the previous section.

\section{Monte Carlo results}

In this section we present the calculations of the spin-stiffnesses for the  
$SO(3)\otimes O(2)$ and $SU(2)\otimes U(1)$ models. Before that, let us first give a 
qualitative visualization of the effect of topology in this problem. 
Figure (\ref{cv}) presents the specific heat as a function of the temperature for 
the two models and for three different lattice sizes: 4$\times$4, 10$\times$10 and 20$\times$20.
Already for these relatively small systems the effect of topology is striking.
In the $SO(3)\otimes O(2)$ case, the specific heat curves show a marked maximum whereas 
it is not the case for the topologically trivial $SU(2)\otimes U(1)$ model. 
This maximum is usually interpreted as the signature of the presence of topological excitations. 
Note that the location of the maximum provides a rough estimate of the cross-over temperature
at which these excitations are activated$^{\cite{kawamura3}}$. 
Here, and without making a detailed analysis based on much bigger sizes, we get approximately $T_V \sim 2.6$. 

\begin{figure}[htp]
\begin{center}
\includegraphics[height=10cm,angle=0]{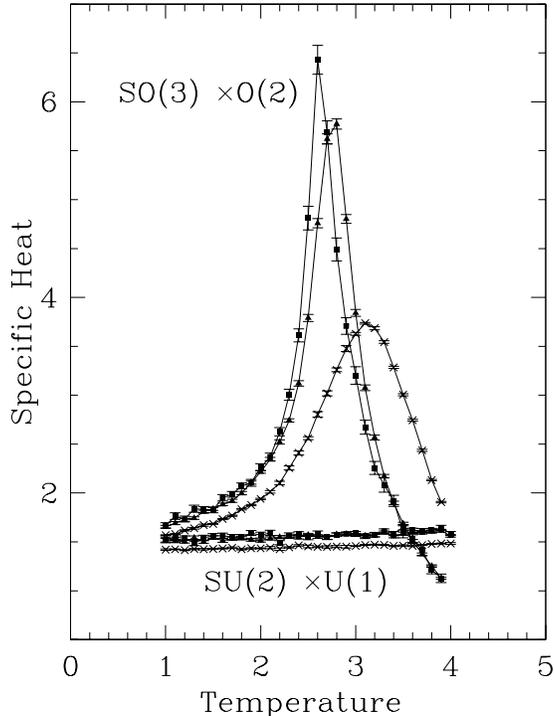}
\caption{Specific heat as a function of the temperature for the
$SO(3) \otimes O(2)$ and $SU(2) \otimes U(1)$ models. Three different sizes have been considered: 
4$\times$4 (crosses), 10$\times$10 (solid triangles), and 20$\times$20 (solid squares). For the $SU(2) \otimes U(1)$
case the curves corresponding to the two largest sizes are undistinguishable. }
\label{cv}
\end{center}
\end{figure}

From our numerical results we propose to distinguish four different temperature regimes. 

\subsection{The spin-wave regime}

The spin-wave regime corresponds to the very-low temperature regime.
We have plotted in Figures (\ref{rhoSO3}) and (\ref{rhoSU2}) the spin-stiffnesses  $\rho_1/T$ and 
$\rho_3/T$ as  functions of $\ln L$ at temperature $T=0.5$
for the $SO(3)\otimes O(2)$ and $SU(2)\otimes U(1)$ models, respectively. The parameters of the action are 
$(p_1,p_1,p_3)=(1,1,0.9)$. The behaviors of the spin-stiffnesses of both models
are in full agreement with the two-loop RG predictions (solid line), Eq.(\ref{cl433}).
As expected from our estimate of $T_V$, these results show that at $T=0.5$ the topological excitations are not yet 
activated and that the physics is controlled by spin-waves well described by the perturbative NL$\sigma$ model. 
For this temperature and for the parameters $p_1$ and $p_3$ chosen, we find an 
almost linear behavior as a function of 
the logarithm of the size. In other words, the two-loop effects are almost negligible. 
The numerical slopes are fully 
compatible with the theoretical slope of ${1/4\pi}$ as given by 
the one-loop equation  (\ref{stiffcorrelation}). Note that, by measuring the two independent 
spin-stiffnesses of the HAFT model, Southern and Young$^{\cite{southern}}$ 
had already confirmed, for one value of the temperature, the RG predictions.

\begin{figure}[htp]
\begin{center}
\includegraphics[height=10cm,angle=0]{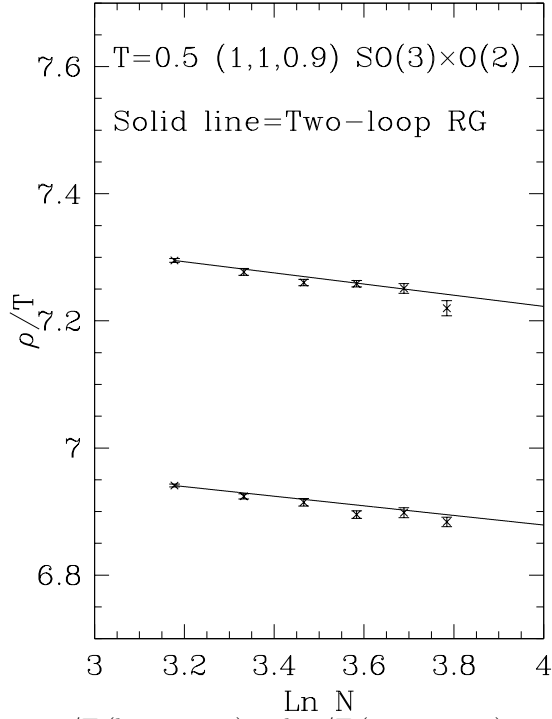}
\caption{$SO(3)\otimes O(2)$ spin-stiffnesses $\rho_1/T$ (lower curve) and $\rho_3/T$ (upper curve) 
as a function of $\ln L$. Temperature $T=0.5$ 
and $(p_1,p_1,p_3)=$(1,1,0.9). The solid line is the two-loop RG prediction as given by Eq.(\ref{cl433}).
Open boundary conditions. Number of clusters used ranges from 5\ $10^6$ to 8\ $10^6$.}
\label{rhoSO3}
\end{center}
\end{figure}

\begin{figure}[htp]
\begin{center}
\includegraphics[height=10cm,angle=0]{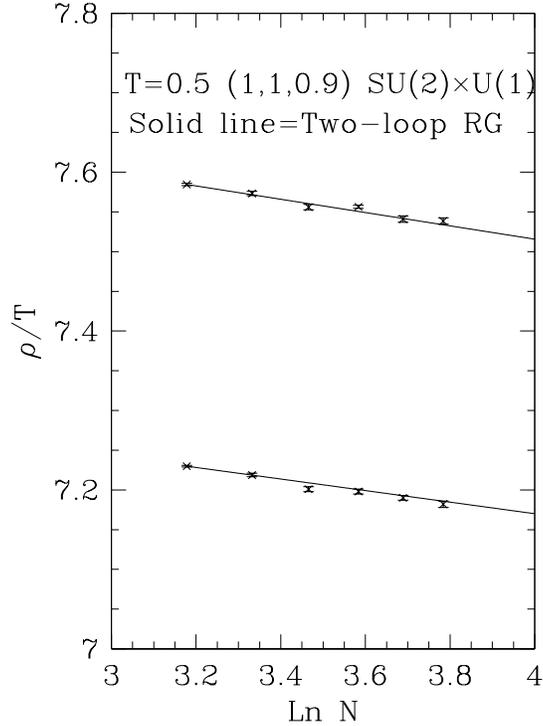}
\caption{$SU(2)\otimes U(1)$ spin-stiffnesses $\rho_1/T$ (lower curve) and $\rho_3/T$ (upper curve) 
as a function of $\ln L$. Temperature $T=0.5$
and $(p_1,p_1,p_3)=$(1,1,0.9). The solid line is the two-loop RG prediction as given by Eq.(\ref{cl433}).
Open boundary conditions. Number of clusters ranges from 2\ $10^6$  to 4\ $10^6$.}
\label{rhoSU2}
\end{center}
\end{figure}

This overall behavior persists up to temperatures of order $T\sim $1 where we enter a new regime. 
Finally, note that the absolute values of the temperature-rescaled spin-stiffnesses for the two models 
at $T=0.5$ are different. Between $T=0$ and 
$T\sim 1$ this difference is almost constant. At $T=0$, this constant 
can be calculated analytically, on the lattice models,  from the  finite parts of
the one-loop counterterms that renormalize the couplings. It is found to be equal to ${5/16}$.

\subsection{The almost-spin-wave regime}
For $T\simeq 1$, the spin-stiffnesses of the two models start to differ: whereas their variations as a 
function of $\ln L$ are correctly described by Eq.(\ref{cl433}) (see Fig.(\ref{stiffrenormalized})) 
the absolute values of $\rho_1/T$ and $\rho_3/T$  for the $SO(3)\otimes O(2)$ model get smaller and smaller compared to 
those of the $SU(2)\otimes U(1)$ model as the temperature increases. 
We have decided to refer to this regime as the almost-spin-wave regime. 

\begin{figure}[htp]
\begin{center}
\includegraphics[height=15cm,angle=0]{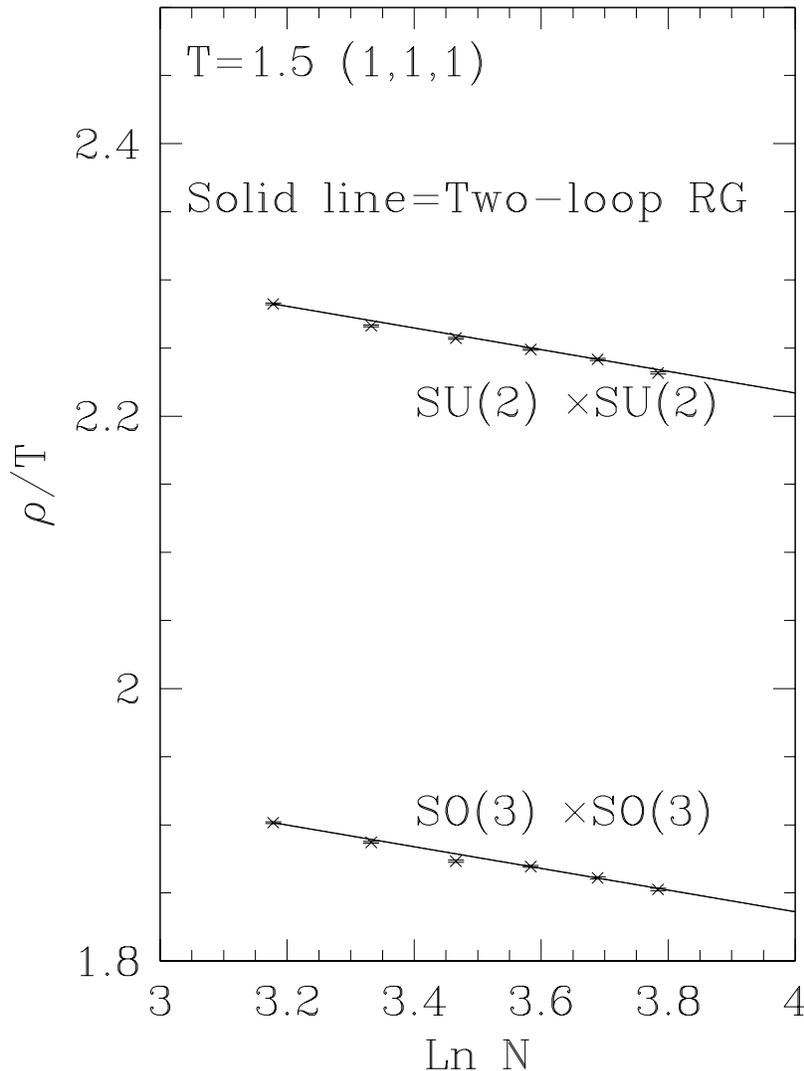}
\caption{Spin-stiffness as a function of $\ln L$ at $T=1.5$ for 
the two models in the symmetric case ($p_1=p_3$).
Note that in this case all three spin-stiffnesses are equal.}
\label{stiffrenormalized}
\end{center}
\end{figure}

Let us now give a quantitative account of this phenomenon. 
Since the difference between $\rho_1/T$ and $\rho_3/T$ turns out to be irrelevant for this discussion, 
we restrict  ourselves from now on to the fully symmetric $SO(3)\otimes SO(3)$ and 
$SU(2)\otimes SU(2)$ models where all three spin-stiffnesses are equal. 
The important point is that in the almost-spin-wave regime, 
the spin-stiffness still displays a linear behavior as the function of $\ln L$ even for the 
topologically non-trivial model. It is thus natural to define a characteristic length $\xi_{eff}$ as:
\be
\rho/T \equiv {1\over 4\pi} \ln {\xi_{eff}\over L}
\label{eq40}
\ee
with $\xi_{eff}$ being a function of $T$ only. We recall that in the topologically trivial 
$SU(2)\otimes SU(2)$ model and at the one-loop approximation, $\xi_{eff}$ is the 
correlation length (see Eq.(\ref{stiffcorrelation})). 
Thus, the simplest hypothesis is that the $\xi_{eff}$ of the $SO(3)\otimes SO(3)$ model 
is still, in this regime,  proportional to the correlation length. 
One can thus expect the ratio: 
\be
R^{MC}(T)\equiv \frac{\xi_{eff}[SU(2) \otimes SU(2)]} {\xi_{eff}[SO(3)\otimes SO(3)]}
\label{rmc}
\ee
to be a good indicator of the influence of the topology. 
We give in Fig.\ref{rapksi}, $R^{MC}(T)$ as a function of the temperature.

\begin{figure}[htp]
\begin{center}
\includegraphics[height=15cm,angle=0]{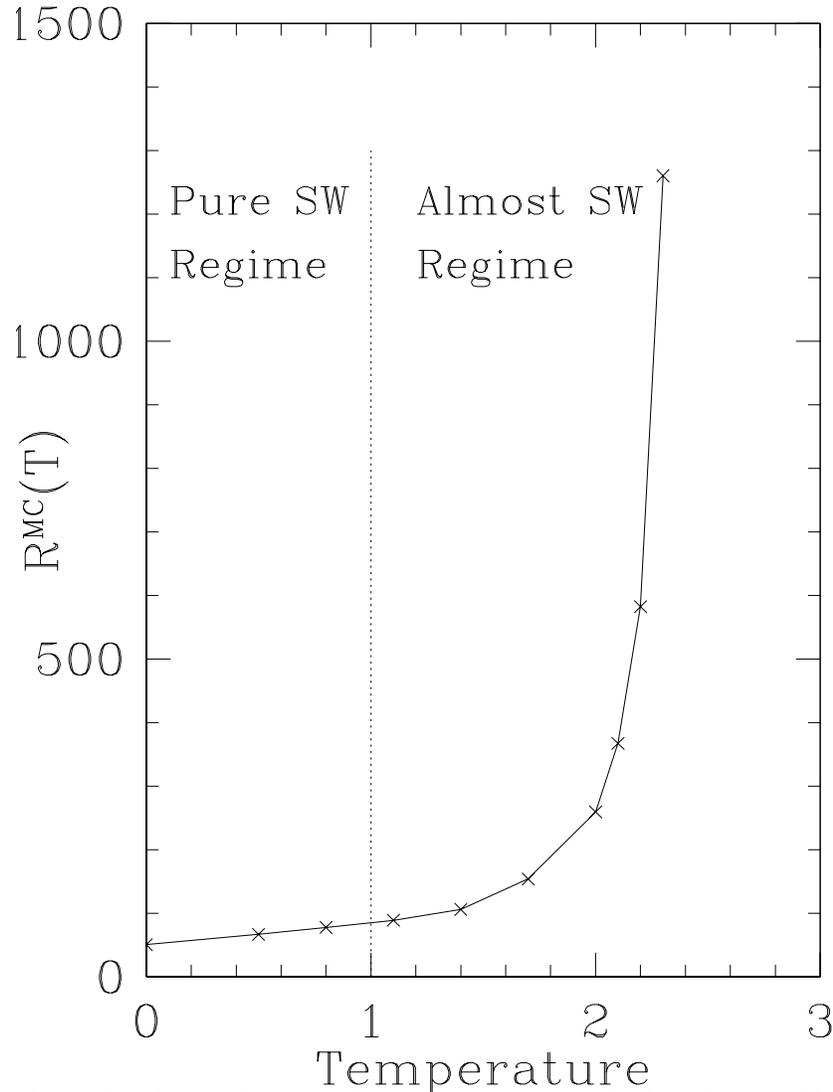}
\caption{Ratio of correlation lengths as a function of the
temperature as defined in the text. The solid line is just to guide the eyes. The separation at $T \sim 1$ between the pure spin-wave (SW) and the almost-spin-wave regimes 
is rather arbitrary.}
\label{rapksi}
\end{center}
\end{figure}

From this figure, we see that at low temperatures (in the spin-wave regime) $R^{MC}(T)$ 
is almost independent on the temperature. 
Note that at $T=0$ it converges to a value different from one because of the constant
shift of $5/16$ between the two temperature-rescaled spin-stiffnesses as discussed above. 
Within the range of temperatures $T\in [1,2.1]$ 
the ratio is found to increase extremely rapidly.  
We have found that the curve can be well fitted using a form:
\be
R^{MC}(T) = C \exp\left[\frac{\alpha}{(T-T_c)^\beta}\right]
\label{rmcfit}
\ee
The ``best'' values found are: $C=23.52,\alpha=1.655,\beta=0.600$, and $T_c= 2.532$. 
This clearly indicates that the topologically non-trivial model becomes more disordered than the 
topologically trivial one at a temperature of order $T_c= 2.5$, 
a value compatible with that obtained from the peak of the specific heat.  Of course, 
the analytical form chosen in Eq.(\ref{rmcfit}) must be taken with lot of caution.  
Many different analytical forms could have been used and give similar results.

\subsection{The vortex regime} 

At temperatures higher than typically $T=2.1$, we enter in a new regime that we propose to call the vortex regime
in which the topological excitations play a major role.
Figures \ref{rhoSO3T2d2}, \ref{rhoSO3T2d4}, \ref{rhoSO3T2d6} and \ref{rhoSU2smallT} 
present, for the two models, the spin-stiffness as a function of the size at the temperatures 
$T=2.2,2.4$ and $2.6$, respectively. 
For the $SU(2)\otimes SU(2)$ model (figure \ref{rhoSU2smallT}) 
the spin-stiffness is still correctly described  by the 
two-loop NL$\sigma$ model predictions, Eq.(\ref{cl433}). 
For the $SO(3)\otimes SO(3)$ model it is still possible to define an effective slope at sufficiently 
small sizes but it now differs significantly from the RG predictions. This effective slope is found to 
increase as a function of the temperature. For example, at $T=2.2$ the slope is about $\sim -0.115$ to 
be compared with $-1/4\pi \sim -0.080$. At $T=2.4$ the slope is $\sim -0.158$, a value approximately two times larger than in the spin-wave regime.
Moreover, the spin-stiffness displays some irregularities which could be associated with 
the presence of long-lived topological configurations that affect the dynamics. 
The most irregular curve has been obtained at temperature $T=2.6$ (figure \ref{rhoSO3T2d6}).

\begin{figure}[htp]
\begin{center}
\includegraphics[height=15cm,angle=0]{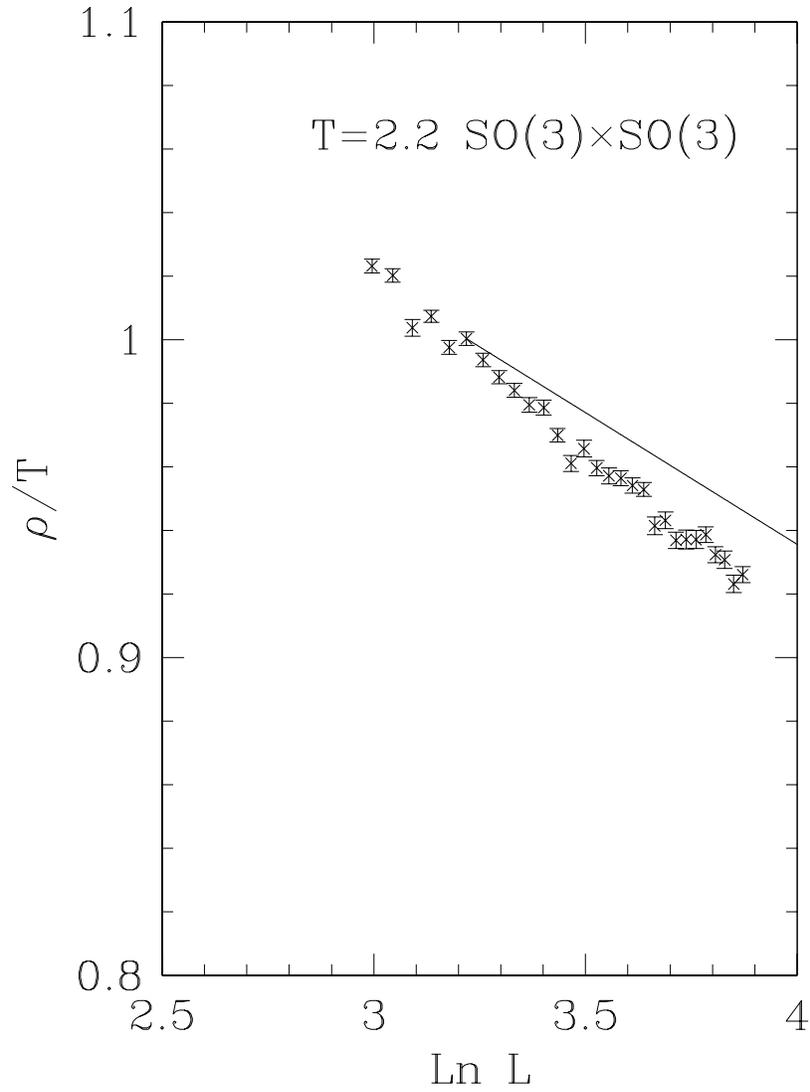}
\caption{$SO(3)\otimes SO(3)$ spin-stiffness as a function of $\ln L$. $T=2.2$.
The solid line is the two-loop RG prediction.}
\label{rhoSO3T2d2}
\end{center}
\end{figure}

\begin{figure}[htp]
\begin{center}
\includegraphics[height=15cm,angle=0]{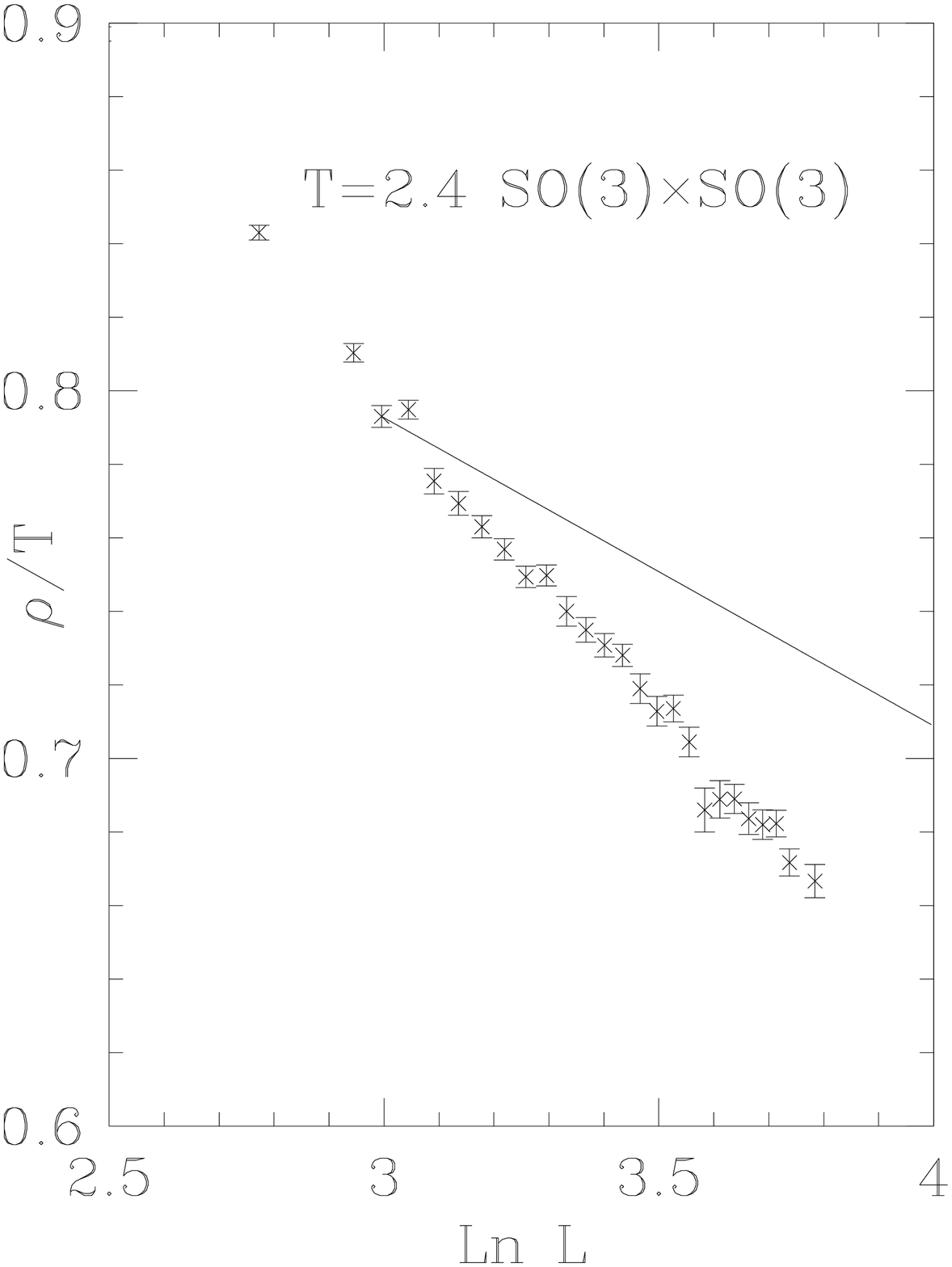}
\caption{$SO(3)\otimes SO(3)$ spin-stiffness as a function of $\ln L$.  $T=2.4$.
The solid line is the two-loop RG prediction.}
\label{rhoSO3T2d4}
\end{center}
\end{figure}

\begin{figure}[htp]
\begin{center}
\includegraphics[height=15cm,angle=0]{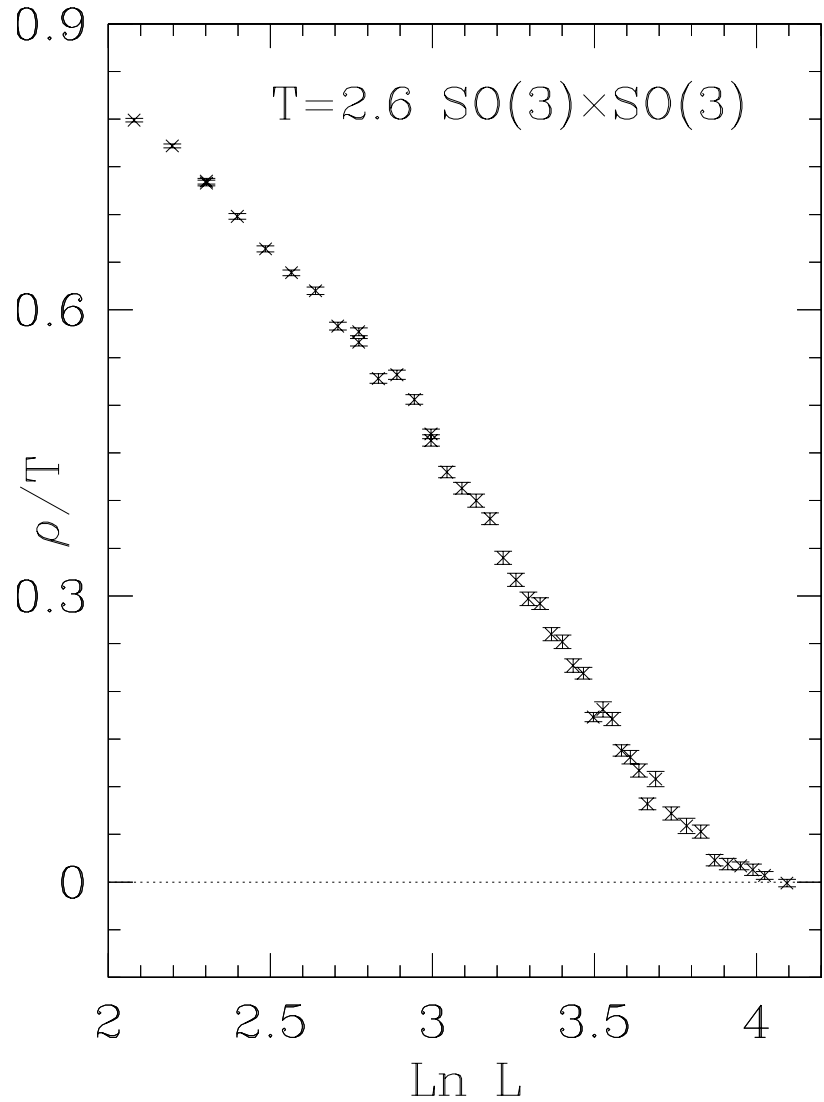}
\caption{$SO(3)\otimes SO(3)$ spin-stiffness as a function of $\ln L$. $T=2.6$} 
\label{rhoSO3T2d6}
\end{center}
\end{figure}

\begin{figure}[htp]
\begin{center}
\includegraphics[height=15cm,angle=0]{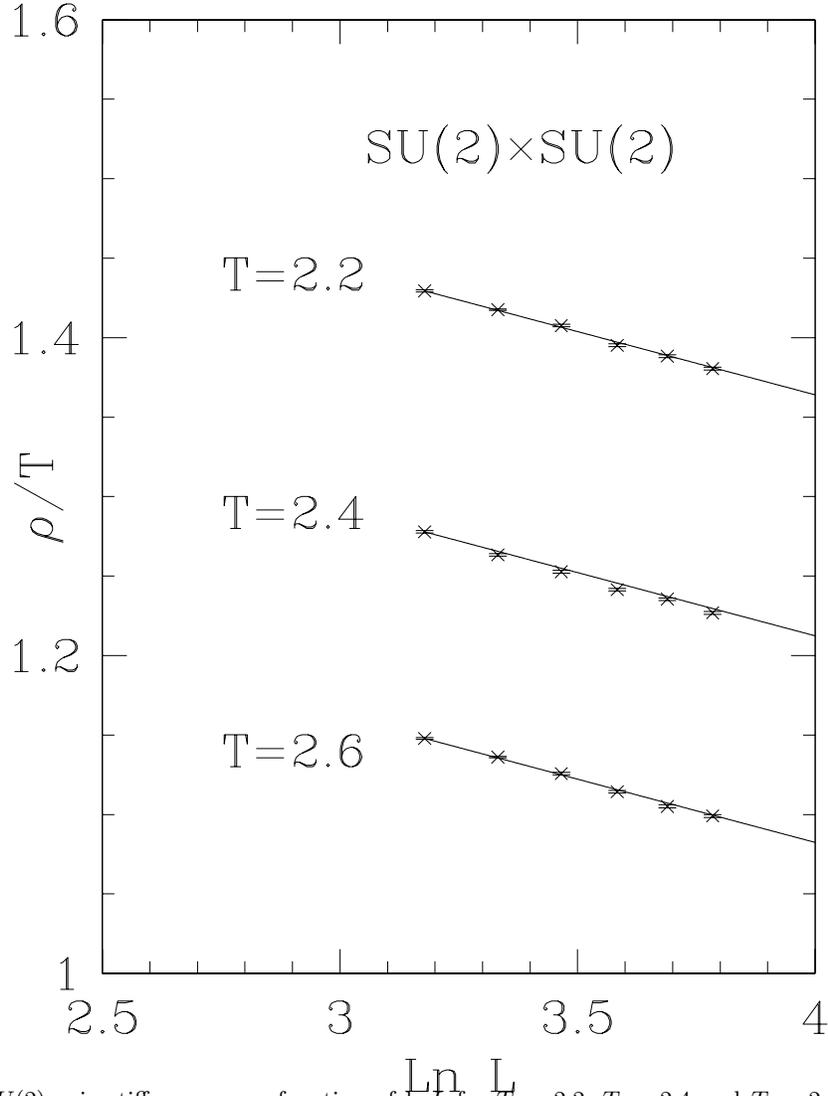}
\caption{$SU(2)\otimes SU(2)$ spin-stiffnesses as a function of $\ln L$ for 
$T=2.2$, $T=2.4$ and $T=2.6$. The solid lines are the two-loop RG predictions.}
\label{rhoSU2smallT}
\end{center}
\end{figure}

\subsection{The high-temperature regime}

At temperatures higher than typically $T=2.6$ the spin-stiffnesses as a function of the size of the 
$SO(3)\otimes SO(3)$ model recovers a smooth behavior.
Fig.(\ref{rhoSO3HT}) presents such a behavior at temperature, $T=2.9$. 
Fig.(\ref{rhoSU2HT}) presents the spin-stiffness for the $SU(2)$ case at $T=7.7$.
The overall behavior of the spin-stiffnesses is quite different.  In the $SO(3)\otimes SO(3)$ case $\rho/T$ 
vanishes abruptly with a change of concavity whereas in the $SU(2)\otimes SU(2)$ case it goes slowly down 
to zero without any change of concavity.

\begin{figure}[htp]
\begin{center}
\includegraphics[height=10cm,angle=0]{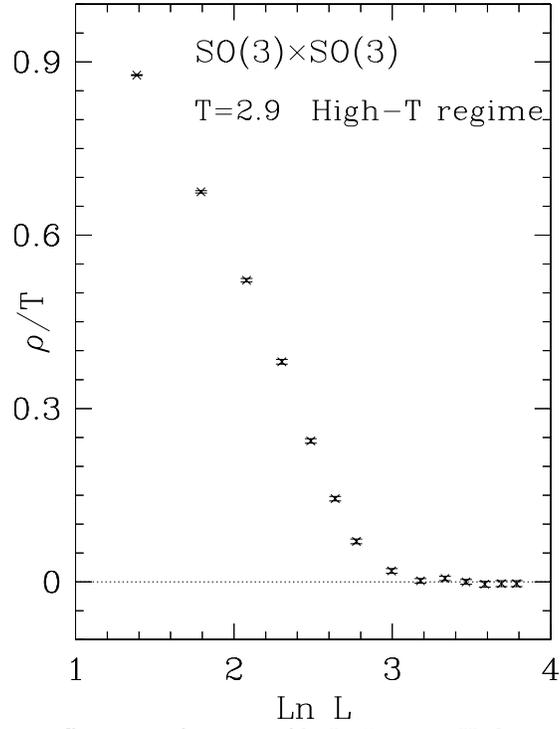}
\caption{$SO(3)\otimes SO(3)$ spin-stiffness as a function of $\ln L$. $T=2.9$. 
High-temperature regime for the model.}
\label{rhoSO3HT}
\end{center}
\end{figure}

\begin{figure}[htp]
\begin{center}
\includegraphics[height=15cm,angle=0]{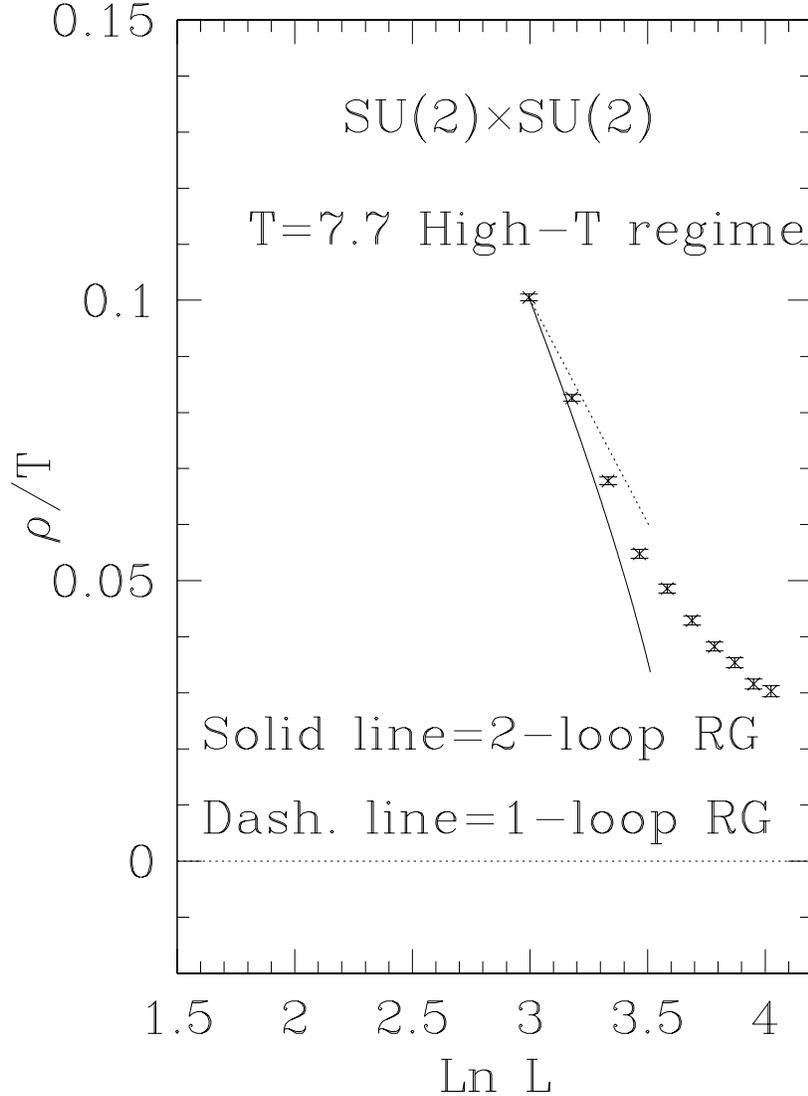}
\caption{$SU(2)\otimes SU(2)$ spin-stiffness as a function of $\ln L$. $T=7.7$. 
High-temperature regime for the model.}
\label{rhoSU2HT}
\end{center}
\end{figure}

\section{Discussion}

We now attempt to give a theoretical analysis of our results. 
It is notoriously difficult to tackle with the physics of $Z_2$ topological defects 
since in this case there is nothing equivalent to the Villain transformation. 
In the usual ferromagnetic case, $O(N)/O(N-1)$, Cardy and Hamber have proposed 
to describe the effect of compacity of the sphere $S_{N-1}$ by means of additional terms 
in the RG equation for the temperature$^{\cite{cardy1}}$. These equations, valid in the 
vicinity of $N=2$ and at order $T^3$ -- which correspond to two-loop in 
perturbation theory -- read: $^{\cite{cardy1}}$
\begin{equation}
\left\{
\begin{array}{l}
\displaystyle{dT(l)\over dl}=(N-2) {T(l)^2\over 2 \pi} + (N-2) {T(l)^3\over (2 \pi)^2} +4\pi^3 y(l)^2+...\\
\\
\displaystyle{dy(l)^2\over dl}=\left(4-{2\pi\over T(l)}\right)y(l)^2+...\\
\end{array}
\right.
\label{cardy}
\end{equation} 
These equations have been derived by assuming analyticity in $y^2$ and in $N$ and by requiring 
that the two following limiting cases are recovered:  
i) the perturbative $\beta$ function of the $O(N)$ model for $N\ge 3$; 
ii) the Kosterlitz-Thouless equations  for $N=2$. 
In fact, Cardy and Hamber have shown that it is the only set of equations compatible with 
these requirements$^{\cite{cardy1}}$.
For $N=2$, one recovers the Kosterlitz-Thouless (K-T) equations$^{\cite{kosterlitz,kosterlitz1}}$ 
where $y$ identifies with the fugacity of vortices. 
The case $y=0$ leads to the two-loop perturbative 
$\beta$ function of the $O(N)/O(N-1)$ NL$\sigma$ model in two dimensions. 
For $N\ne 2$, $y$ lacks of a clear interpretation but it has been conjectured that 
it encodes the effect of compacity. 
Clearly, the ($N>3$, $y\ne 0$)-case we consider here could very well lie
outside the domain of validity of these equations. It is therefore important to 
insist on the fact that they must only be considered as some phenomenological 
RG equations for a model displaying topological defects, $y$ playing the role of a fugacity by 
analogy with the K-T case.
Our aim is to show that they are able to reproduce the gross features of 
the behavior of the spin-stiffness and correlation length found in our case. 
As in the K-T case, the physical ``fugacity'' $y(l=0)$ that appears as the initial 
condition in Eq.(\ref{cardy}) is not independent on the temperature. 
However, in contrast with this last case its dependence on the physical 
temperature $T(l=0)$ is unknown. The simplest assumption we can think of is:
\begin{equation}
y(l=0)=e^{-\gamma/T(l=0)}
\label{fug}
\end{equation}
as in the K-T case. In Eq.(\ref{fug}), $\gamma$ is an adjustable parameter. 

The equations (\ref{cardy}) are considered in the case $N=4$ since $SO(3)\otimes SO(3)/SO(3)$ and $SO(4)/SO(3)$ have the same spin-wave content and differ by their topological properties. These latter properties
are expected to be taken into account via the $y$-terms in Eq.(\ref{cardy}).
Moreover, we write them in terms of the spin-stiffness of the $SO(3)\otimes SO(3)/SO(3)$ model. 
These equations are obtained by the substitution $1/T(l)\to 4\widetilde{\rho}(l)$:}
\begin{equation}
\left\{
\begin{array}{l}
\displaystyle{d\widetilde{\rho}(l)\over dl}=-{1\over 4 \pi} -{1\over 32 \pi^2 \widetilde{\rho}(l)}-16\pi^3 y(l)^2 \widetilde{\rho}(l)^2\\
\\
\displaystyle{dy(l)^2\over dl}=\left(4-{8\pi\over \widetilde{\rho}(l)}\right)y(l)^2.\\
\end{array}
\right.
\label{cardysimple}
\end{equation}
By direct integration of Eqs.(\ref{cardysimple}) 
up to the scale $L$ of the lattice size -- $l=\ln L/a$ -- one can obtain the dependence 
of $\widetilde{\rho}(l)$  on the temperature $T(l=0)$ and on $L$.
To make contact with our Monte Carlo results, we are also interested in the  correlation
length $\xi$. This last quantity is defined, as usual, from the fact that when the RG 
scale $e^l$ becomes of the order of $\xi$, the spin-stiffness $\widetilde{\rho}(l)$
vanishes, see Eq.(\ref{eq40}):
\begin{equation}
\xi/a \sim e^l\ \ \ \ {\hbox{with}}\ \ \ \ \hspace{0.5cm} \widetilde{\rho}(l)\simeq 0\ .
\label{longueurdecorrelation}
\end{equation}
As in the Monte Carlo simulations, 
we compute from Eqs.(\ref{cardysimple}) the spin-stiffnesses and correlation 
lengths in both situations: without and with defects. This consists in setting 
respectively $y(l)$ to zero or not in Eq.(\ref{cardysimple}). 
Note finally that, for simplicity, we have chosen to take the same normalization at 
zero-temperature for the spin-stiffnesses with and without defects.

Let us now show that we retrieve the essential features of the regimes previously 
identified except, obviously, for the high-temperature regime which is out of reach of 
the RG equations (\ref{cardysimple}) which, for the spin-wave part, 
are perturbative in the temperature. 
At very low temperatures, $y(l)$ is very small and remains small along the RG flow. 
As a result we find almost no difference between the systems with and without vortices: 
this is the spin-wave regime that we recover trivially. 

\begin{figure}[htp]
\begin{center}
\includegraphics[height=15cm,angle=0]{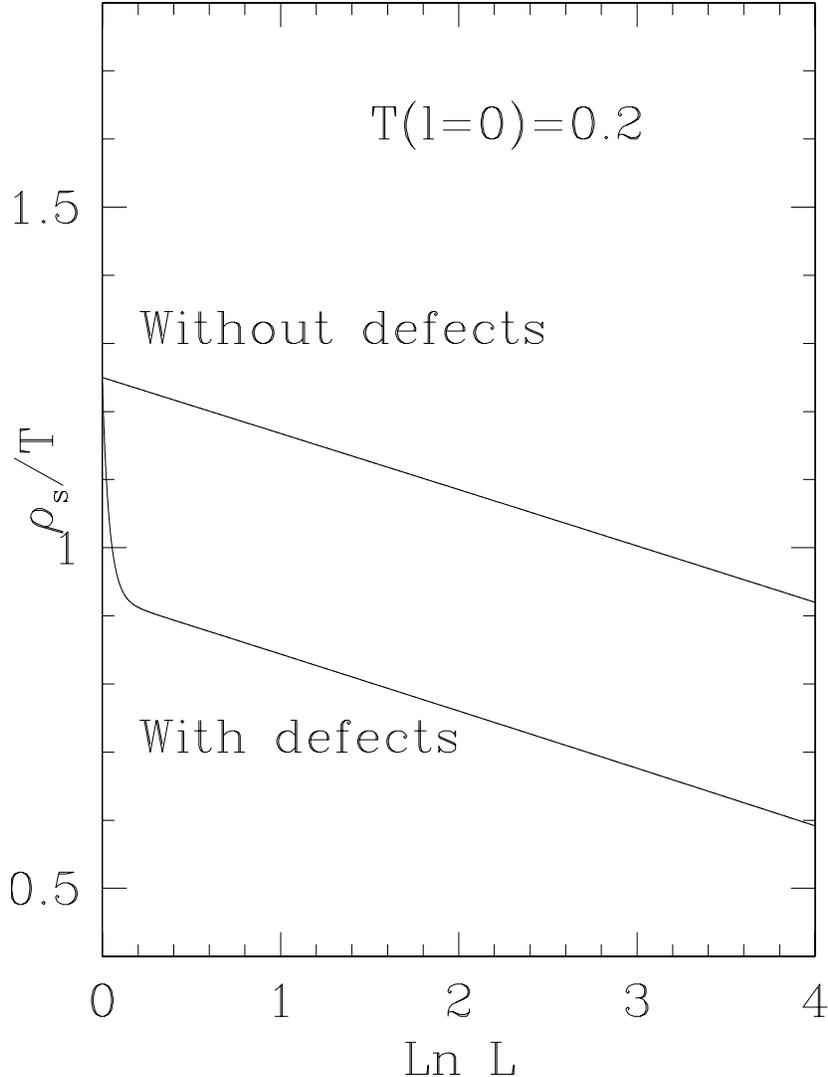}
\caption{Spin-stiffness as a function of $\ln L$ with and without 
defects. $T(l=0)=0.2.$}
\label{rhofonctiondelcardy}
\end{center}
\end{figure}

As the temperature increases, the $y$-term plays a more and more important role. 
Adjusting the free parameter $\gamma$ at a value $\gamma=0.45$ 
the temperature where the defects start to play a significant role is typically 
$T(l=0)\sim 0.15$. We plot in Fig.(\ref{rhofonctiondelcardy}) 
the spin-stiffnesses as a function of 
the system size at a slightly higher temperature, $T(l=0)=0.2$. The upper curve 
corresponds to the system without defects and the lower one to that with defects. 

As in our Monte Carlo simulations, we find that the absolute value of the spin-stiffness 
is decreased by the presence of defects.  After an abrupt jump at very small sizes the spin-stiffness is
found to be linear as a function $\ln L$. 
Up to an accuracy of a few percent, the slope is not affected by the defects. Its value, 
$-0.080 \sim -1/4\pi$, corresponds to the perturbative RG result.
This behavior is similar to that predicted  by the spin-wave analysis, except that the
absolute value of the spin-stiffness is smaller.
This corresponds to the almost-spin-wave regime previously identified.  

\begin{figure}[htp]
\begin{center}
\includegraphics[height=15cm,angle=0]{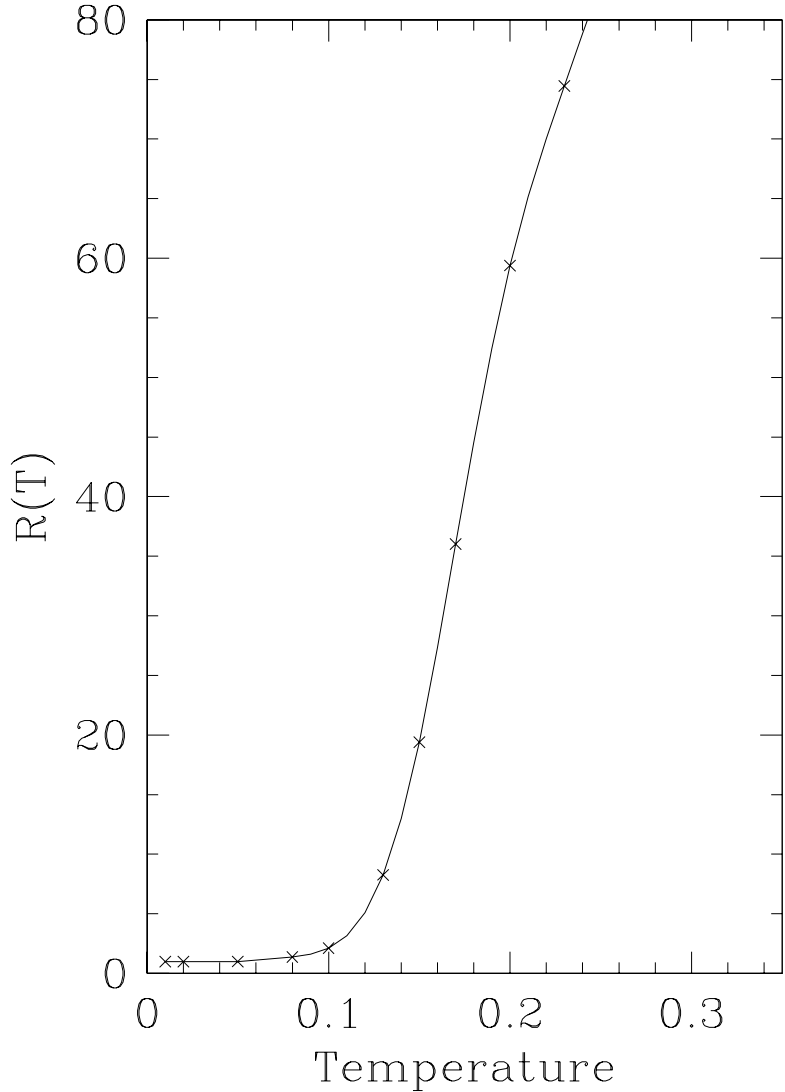}
\caption{Ratio $R(T)$ of correlation lengths as a function of $T(l=0)$.}
\label{ratioxicardy}
\end{center}
\end{figure}

To obtain a completely consistent picture, 
it is also necessary to see whether the ratio of the correlation lengths, 
without and with defects, considered as a function of $T(l=0)$:
\begin{equation}
R(T)={\xi_{y(l)=0}\over \xi_{y(l)\ne 0}}
\end{equation}
behaves as $R^{MC}$, Fig.(\ref{rapksi}). We give in Fig.(\ref{ratioxicardy}), 
the ratio $R$ obtained by direct integration of Eq.(\ref{cardysimple}).
This figure clearly shows a behavior comparable to that observed in Fig.({\ref{rapksi}}). 

\begin{figure}[htp]
\begin{center}
\includegraphics[height=15cm,angle=0]{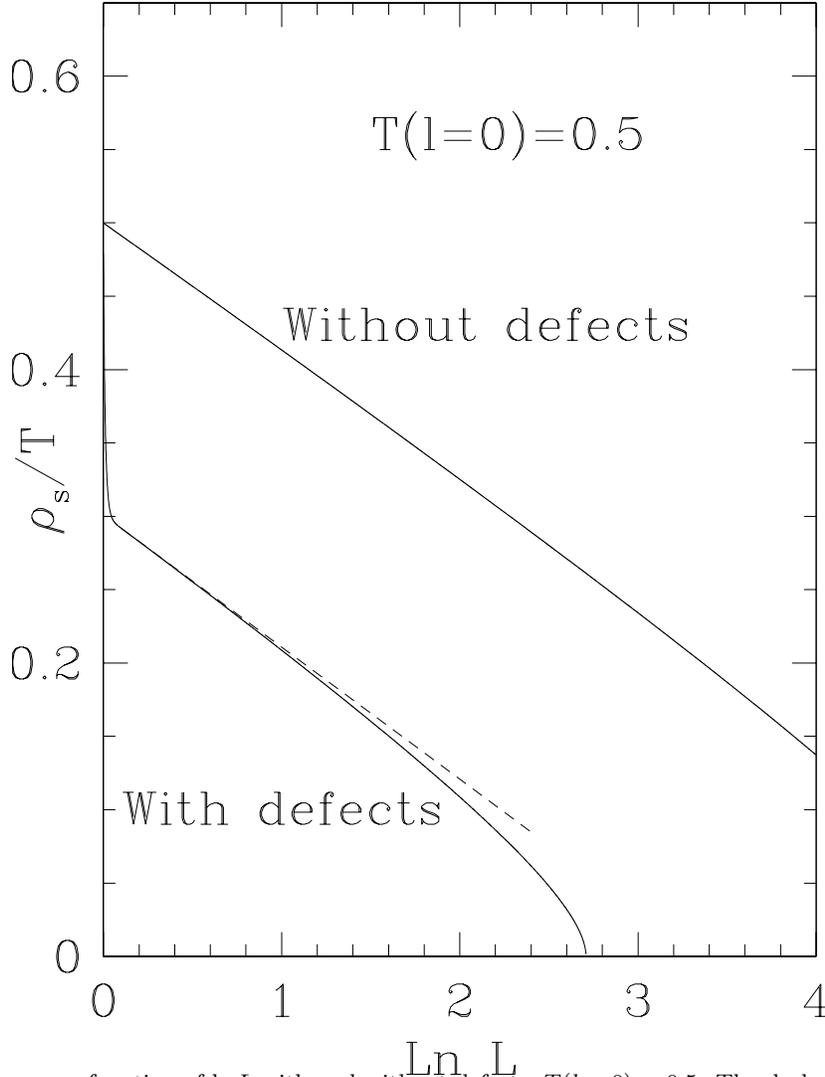}
\caption{Spin-stiffness as a function of $\ln L$ with and without defects,
$T(l=0)=0.5$. The dashed line represents 
the linear behavior associated with the model without defects. A slight increase of the slope is 
observed when the defects are present.}
\label{rhofonctiondelcardy2}
\end{center}
\end{figure}

At higher temperature, one enters in a regime where the spin-stiffness 
as a function of the size begins to display a different behavior. 
Figure (\ref{rhofonctiondelcardy2}) presents the spin-stiffnesses at $T=0.5$ (to facilitate 
the comparaison the dashed line represents the linear behavior associated with the model
without defects). 
It is  possible to define a linear regime but now with a slightly greater slope 
than in the defect-free case. 
The difference between the two slopes is about 10$\%$. 
As the temperature is increased the spin-stiffness still displays a 
linear behavior but now with an increasing temperature-dependent slope.

Such a behavior has been numerically observed in the vortex regime (see, Fig.(\ref{rhoSO3T2d2}) 
and (\ref{rhoSO3T2d4}) where some ``effective'' linear behavior with a larger slope is observed).
However, note that the irregular behavior found in Fig.(\ref{rhoSO3T2d6}) is not reproduced here. 
Of course, the range of temperature over which this linear behavior is observed, 
as well as the variation of the slope with $T(l=0)$ 
depends rather strongly in our calculation on the relation 
between $y(l=0)$ and $T(l=0)$, Eq.(\ref{fug}). With
our choice of $\gamma$, the maximum variation found for the slope as a function of the 
temperature is about 15$\%$ which is somewhat below what is obtained in the simulations. 
This could certainly be corrected by another choice of $y(T)$.

At even higher temperature, Eqs.(\ref{cardysimple}) are no longer valid since they are based on a 
low-temperature expansion and the comparison with the numerical results does not make sense.

The preceding analysis shows that the simple set of equations (\ref{cardysimple}) 
together with the relation (\ref{fug}) seem to capture 
 some of the important features of the presence 
of topological defects in the almost-spin-wave and vortex regimes. Of course, 
only a microscopical approach of the problem could allow to go beyond this semi-quantitative description.

B.D. and D.M. thank B. Dou\c{c}ot and J. Vidal for discussions. 
LCT, LPTL, and LPTHE are Laboratoires associ\'es au CNRS: UMR 7676, 7600, 
and 7589.

E-mail: michel.caffarel@lct.jussieu.fr, azaria@lptl.jussieu.fr, 
delamotte@lpthe.jussieu.fr, and mouhanna@lpthe.jussieu.fr

\end{document}